\documentclass[sn-nature]{sn-jnl}% Style for submissions to Nature Portfolio journals
\usepackage{graphicx}%
\usepackage{multirow}%
\usepackage{amsmath,amssymb,amsfonts}%
\usepackage{amsthm}%
\usepackage{mathrsfs}%
\usepackage[title]{appendix}%
\usepackage{xcolor}%
\usepackage{textcomp}%
\usepackage{manyfoot}%
\usepackage{booktabs}%
\usepackage{algorithm}%
\usepackage{algorithmicx}%
\usepackage{algpseudocode}%
\usepackage{listings}%

\begin{document}

\title[Article Title]{Functional Assessment of Cerebral Capillaries using Single Capillary Reporters in Ultrasound Localization Microscopy}

%%=============================================================%%
%% Prefix	-> \pfx{Dr}
%% GivenName	-> \fnm{Joergen W.}
%% Particle	-> \spfx{van der} -> surname prefix
%% FamilyName	-> \sur{Ploeg}
%% Suffix	-> \sfx{IV}
%% NatureName	-> \tanm{Poet Laureate} -> Title after name
%% Degrees	-> \dgr{MSc, PhD}
%% \author*[1,2]{\pfx{Dr} \fnm{Joergen W.} \spfx{van der} \sur{Ploeg} \sfx{IV} \tanm{Poet Laureate} 
%%                 \dgr{MSc, PhD}}\email{iauthor@gmail.com}
%%=============================================================%%

\author*[1]{\fnm{Stephen} \sur{Lee}}\email{stephen.lee@polymtl.ca}
\author[1]{\fnm{Alexis} \sur{Leconte}}
\author[1]{\fnm{Alice} \sur{Wu}}
\author[2]{\fnm{Joshua} \sur{Kinugasa}}
\author[1]{\fnm{Jonathan} \sur{Porée}}
\author[3]{\fnm{Andreas} \sur{Linninger}}
\author*[1,4]{\fnm{Jean} \sur{Provost}}\email{jean.provost@polymtl.ca}

\affil*[1]{\orgdiv{Department of Engineering Physics}, \orgname{Polytechnic Montreal}, \orgaddress{\street{2500 Chemin de Polytechnique}, \city{Montreal}, \postcode{H3T 1J4}, \state{QC}, \country{CA}}}

\affil[2]{\orgdiv{Department of Biomedical Engineering}, \orgname{Chiba University}, \orgaddress{\street{1-33 Yayoicho}, \city{Chiba}, \postcode{263-8522}, \country{JP}}}

\affil[3]{\orgdiv{Department of Biomedical Engineering}, \orgname{University of Illinois Chicago}, \orgaddress{\street{1200 W Harrison St}, \city{Chicago}, \postcode{60607}, \state{IL}, \country{USA}}}

\affil[4]{\orgname{Montreal Heart Institute}, \orgaddress{\street{5000 Rue Belanger}, \city{Montreal}, \postcode{H1T 1C8}, \state{QC}, \country{CA}}}

%%==================================%%
%% sample for unstructured abstract %%
%%==================================%%

\abstract{The brain's microvascular cerebral capillary network plays a vital role in maintaining neuronal health, yet capillary dynamics are still not well understood due to limitations in existing imaging techniques. Here, we present Single Capillary Reporters (SCaRe) for transcranial Ultrasound Localization Microscopy (ULM), a novel approach enabling non-invasive, whole-brain mapping of single capillaries and estimates of their transit-time as a neurovascular biomarker. We accomplish this first through computational Monte Carlo and ultrasound simulations of microbubbles flowing through a fully-connected capillary network. We unveil distinct capillary flow behaviors which informs methodological changes to ULM acquisitions to better capture capillaries \textit{in vivo}. Subsequently, applying SCaRe-ULM \textit{in vivo}, we achieve unprecedented visualization of single capillary tracks across brain regions, analysis of layer-specific capillary heterogeneous transit times (CHT), and characterization of whole microbubble trajectories from arterioles to venules. Lastly, we evaluate capillary biomarkers using injected lipopolysaccharide to induce systemic neuroinflammation and track the increase in SCaRe-ULM CHT, demonstrating the capability to detect subtle capillary functional changes. SCaRe-ULM represents a significant advance in studying microvascular dynamics, offering novel avenues for investigating capillary patterns in neurological disorders and potential diagnostic applications.}


%%\pacs[JEL Classification]{D8, H51}

%%\pacs[MSC Classification]{35A01, 65L10, 65L12, 65L20, 65L70}

\maketitle

\section{Introduction}\label{sec1}
The microvascular system within the brain - comprising a complex network of blood vessels including arteries, arterioles, capillaries, venules, and veins - plays a fundamental role in maintaining the homeostasis and integrity of the central nervous system (CNS) \cite{kamouchi2011brain}. Neuronal health is intimately linked to the function of the microvasculature \cite{dalkara2015cerebral}. Emerging evidence suggests that disruptions in microvascular function are associated with the onset, maintenance, and progression of neurodegenerative conditions \cite{dalkara2011brain}. Notably, at the capillary level, where oxygen exchange occurs and red blood cell (RBC) velocity is at its lowest, is of particular importance since anomalies in capillary flow dynamics, such as turbulent flow, endothelial damage \cite{huang2014vivo}, or intermittent RBC stalling \cite{crumpler2021capillary}, have been correlated with cognitive decline and worsening neurodegeneration \cite{sweeney2018role}. Moreover, disruptions at the single capillary level can cascade downstream, impacting the entire microvasculature and neural networks \cite{zhu2023single}. Given that every capillary is responsible for the supply and maintenance of clusters of neurons \cite{zlokovic2005neurovascular}, there is a critical need for advanced \textit{in vivo} imaging techniques capable of precisely assessing single capillary function throughout the whole brain, advancing neuroimaging capabilities and diagnosis of CNS pathologies.

While there exist microvascular imaging methods, such as optical clearing, micro-CT, or two-photon optical imaging \cite{zhu2021tissue}, that can either achieve high spatial resolutions, temporal fidelity, or whole-brain penetration, there does not exist one that achieves all criteria \textit{in vivo} at the capillary level. One promising approach to whole-brain microvascular imaging is ultrasound localization microscopy (ULM), which has the unique capability to provide non-invasive, deep-tissue imaging of microvascular morphology \cite{couture2018ultrasound,shin2024context}. By leveraging the circulation of microbubbles within the vasculature, ULM enables the reconstruction of super-resolution images detailing micron-level vessel structures and blood flow velocities \cite{couture2018ultrasound,heiles2022performance}. Though ULM has surpassed the diffraction limit in ultrasound imaging, extracting functional information remains an active area of research. Dynamic ULM (DULM), aims to extract additional functional information beyond structural imaging or flow velocity, such as pulsatility \cite{bourquin2021vivo,bourquin2024quantitative}, through synchronization with the cardiac cycle. Functional ULM aims to image the neurovascular response to somatosensory stimuli \cite{renaudin2022functional}. Moreover, a study by \citep{denis2023sensing,chabouh2024whole} demonstrated the detection of renal glomeruli based on microbubble rotational behavior, showcasing the potential for innovative functional applications accessible through ULM. Yet, crucial barriers still exist in achieving single capillary resolutions such as the dependency on multiple microbubble track aggregates to confirm the presence of a vessel, as well as the likely filtration of extremely low capillary velocities that overlap with tissue and skull movement.

In this study, we present single capillary reporters (SCaRe) ULM, capable of non-invasively imaging and measuring capillary function throughout the entire brain at the single-capillary level \textit{in vivo}, at a scale which has not been achieved before. Our approach was developed first via several layers of computational modeling of microbubble flow within realistic simulations of the mouse brain microvasculature \cite{0271678X16671146,0271678X231214840} with fully-connected closures of millions of vessels, built upon state-of-the-art 3D optical imaging and segmentation \cite{linninger2019mathematical,belgharbi2023anatomically}. This \textit{in silico} model allowed us to anticipate the behaviors of microbubbles within capillaries and to propose new strategies for overcoming existing limitations of exiting ULM techniques in capturing slow capillary flow. Specifically, SCaRe is comprised of 1) long ensemble singular value decomposition (LE-SVD), 2) continuous ULM acquisition, and 3) trained hidden Markov models (HMMs). This methodology achieves novel measurement of capillary heterogeneous transit time (CHT) and the mapping of microbubble trajectories based solely on their behavior as they enter and exit capillaries throughout the whole brain. Taken together, SCaRe-ULM unveils spatiotemporal patterns of capillary function, potentially providing new insights into neural pathologies, contributing to the development of early diagnostics in neurodegenerative conditions or new therapeutic interventions.

\section{Results}\label{sec2}

Our understanding of microbubble behavior within capillaries remains limited, except for their sparsity and estimated perfusion times \cite{christensen2019poisson,lowerison2020vivo}. Moreover, tracking a single microbubble poses a challenge when no other tracks are present to facilitate delineation of a vessel. i.e., how can we know whether a single microbubble track is contained by a vessel? Given the rare occurrence of multiple microbubbles traversing the same capillary within a short time frame \cite{hingot2019microvascular}, sophisticated methods for vessel reconstruction and functional measurement are imperative. Yet, comprehensive knowledge regarding the expected behavior of a microbubble bolus injection in the mouse brain remains elusive. Without insights into the ground truth velocities and positions of microbubbles throughout the brain, the development of ULM-based strategies for capillary flow recovery remains inherently challenging. Thus, we sought to investigate whether realistic computational models of the hemodynamic microvascular network could unveil distinct microbubble behaviors within ULM. Specifically, we aimed to address two key questions: 1) How do microbubbles traverse the vascular tree from arteries to veins in ULM? and 2) What changes to ULM can be made to increase the likelihood of recovering single capillary tracks?

\subsection{Modeling Microbubble Flow in Realistic Mouse Microvasculatures}\label{subsec1}

\begin{figure}
    \centering
    \includegraphics[width=0.98\textwidth]{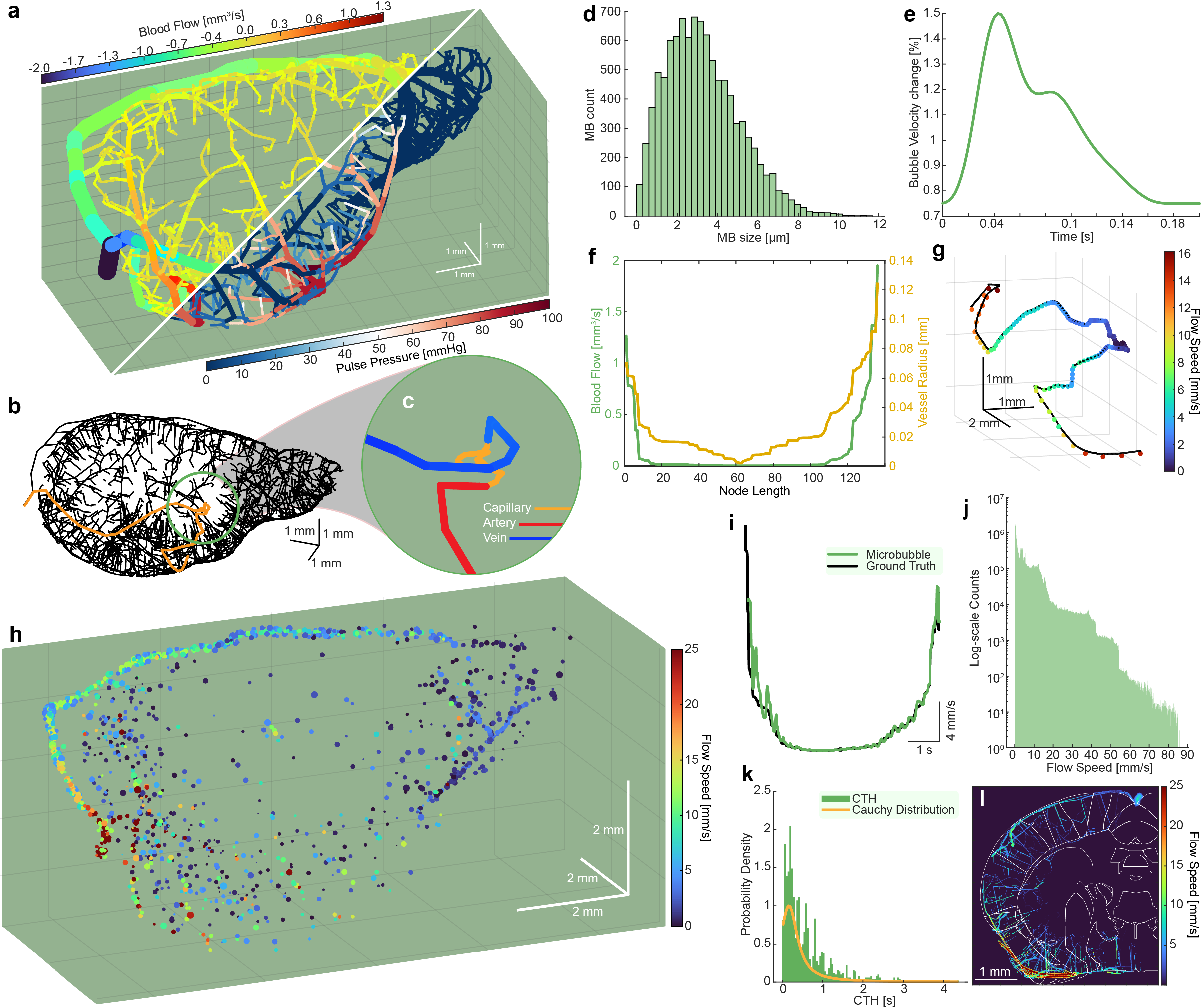}
    \caption{\textbf{a.} Synthetic mouse brain microvascular network with fully-connected capillary networks, blood flow (top), vessel diameter, and pulse pressure (bottom). \textbf{b.} Skeletonized vascular network with a single full vascular path identified. \textbf{c.} Zoomed color-coded image of a single vascular. \textbf{d.} Simulated Definity microbubble distribution from which simulated results are sampled. \textbf{e.} Sample pulsatile waveform calculated using pulse decomposition analysis. \textbf{f.} Sample blood flow and microvessel radius of the path depicted in \textbf{b-c} used to forward propagate simulated microbubbles. \textbf{g.} Resultant microbubble trajectory along path depicted in \textbf{b-c} using initial conditions in \textbf{d-f}. \textbf{h.} Representative fully populated microbubble dataset through the synthetic mouse brain (see Supplemental Video 1). \textbf{i.} Distinct U-shaped velocity profile of the path depicted in \textbf{b-c,g}. \textbf{j.} Log histogram of sampled microbubble velocities throughout the whole dataset in \textbf{h}. \textbf{k.} Sampled capillary heterogeneous transit-times, measured as time required for a microbubble to traverse a capillary network, fitted with a Cauchy distribution. \textbf{l.} Ground truth microbubble paths with the Allen Institute mouse brain atlas overlaid on top, showing regional differences in microbubble flow speed.}
    \label{fig:1}
\end{figure}

Among the various computational models of microvascular flow in the rodent brain, we focused on synthetic networks based on state-of-the-art neuroimaging data (Figure \ref{fig:1}a). These networks emulate complete and balanced circulation, encompassing fully-connected capillary closures and topology from arteries to veins (for a full description of the image-based synthesis and hemodynamic simulation of full mouse brains see \cite{hartung2021mathematical,linninger2019mathematical}). Our hemodynamic simulations can predict oscillatory blood pressure and flow rates at nodes and edges of directed graphs using a graph theoretical approach developed for large scale vascular anatomical networks \cite{pcbi.1008584,cnm.3288}. By constructing a dataset of microbubbles flowing through the microvasculature, we can retrieve all feasible paths from inlets (arteries) to outlets (veins), traversing a single capillary (Figure \ref{fig:1}b-c). This enables precise quantification of movement, microbubble velocity, and transit time behavior throughout the capillary network, given blood flow, vessel radius, and pulse pressure as computed with methods described in \cite{0271678X231214840} and \cite{pcbi.1006549}.

To generate synthetic microbubble ultrasound data, we used sequential Monte Carlo simulations, randomly sampling specific paths (Figure \ref{fig:1}b-c) given initial conditions (Figure \ref{fig:1}d-f)). Using a mouse brain hemisphere, this approach yielded a total of 291,372 potential paths (876,161 nodes; 1,167,529 edges) for microbubble flow (Supplemental Video 1) and for each, we sampled from a microbubble distribution, modeled as a polydispersed-sized bolus of Definity microbubbles \cite{talu2007tailoring} (Figure \ref{fig:1}d). Subsequently, we calculated the associated pulsatile waveform via pulse decomposition analysis \cite{baruch2011pulse}, applying a 75\% variation from peak systole to peak diastole in blood velocity for all vessels (Figure \ref{fig:1}e). Radially distributed blood flow velocities for each vessel segment and instance in time were computed from the graph-based volumetric blood flow rates under parabolic profile assumptions \cite{cnm.3288}. Thus, instantaneous flow speed and vessel diameters were retrieved (Figure \ref{fig:1}f) and used to calculate bubble velocity and radial distance from the centerline along the vessel path \cite{belgharbi2023anatomically}. Finally, we forward propagated the selected microbubbles from inlet to outlet based on the initial conditions and traveling pulse wave velocity \cite{marshall2023alterations,janssen2016need}. Simulation parameters can be found in supplementary table 1.

Figure \ref{fig:1}h illustrates a fully populated mouse brain microvascular network following population with randomly sampled microbubble paths (Supplementary Video 1). The microbubbles exhibit synchronized pulsatile waveforms propagating through the vascular tree. Thus, we obtain multiple "ground truth" datasets comprising microbubble position, size, and instantaneous flow velocity in mm/s within the 3D mouse brain. The key insight here is the discovery of distinct U-shaped velocity behaviors as a single microbubble traverses the vascular network with low velocities (\(\textless\) 5 mm/s) at the capillary mesh (Figure \ref{fig:1}i). In this dataset, microbubble velocities range from 0 to 85 mm/s (Figure \ref{fig:1}j). Furthermore, we can estimate the CHT within the whole vascular network, measured as the time it takes for a bubble to move from the start to the end of a capillary (areas of velocity \(\textless\) 2 mm/s) (Figure \ref{fig:1}k). We find that the estimated CHT appears to conform to a Cauchy Distribution. Crucially, we see that CTH can be as long as 4 seconds. Furthermore, we can measure regional CTH (Figure \ref{fig:1}l) and start identifying spatiotemporal patterns. Taken together, we identify that a factor preventing capillary imaging is the time required to track the complete microbubble behavior, necessitating either continuous acquisition or long acquisition times.

\subsection{Sequential Monte Carlo simulations reveal distinct capillary characteristics}\label{subsec2}

\begin{figure}
    \centering
    \includegraphics[width=0.98\textwidth]{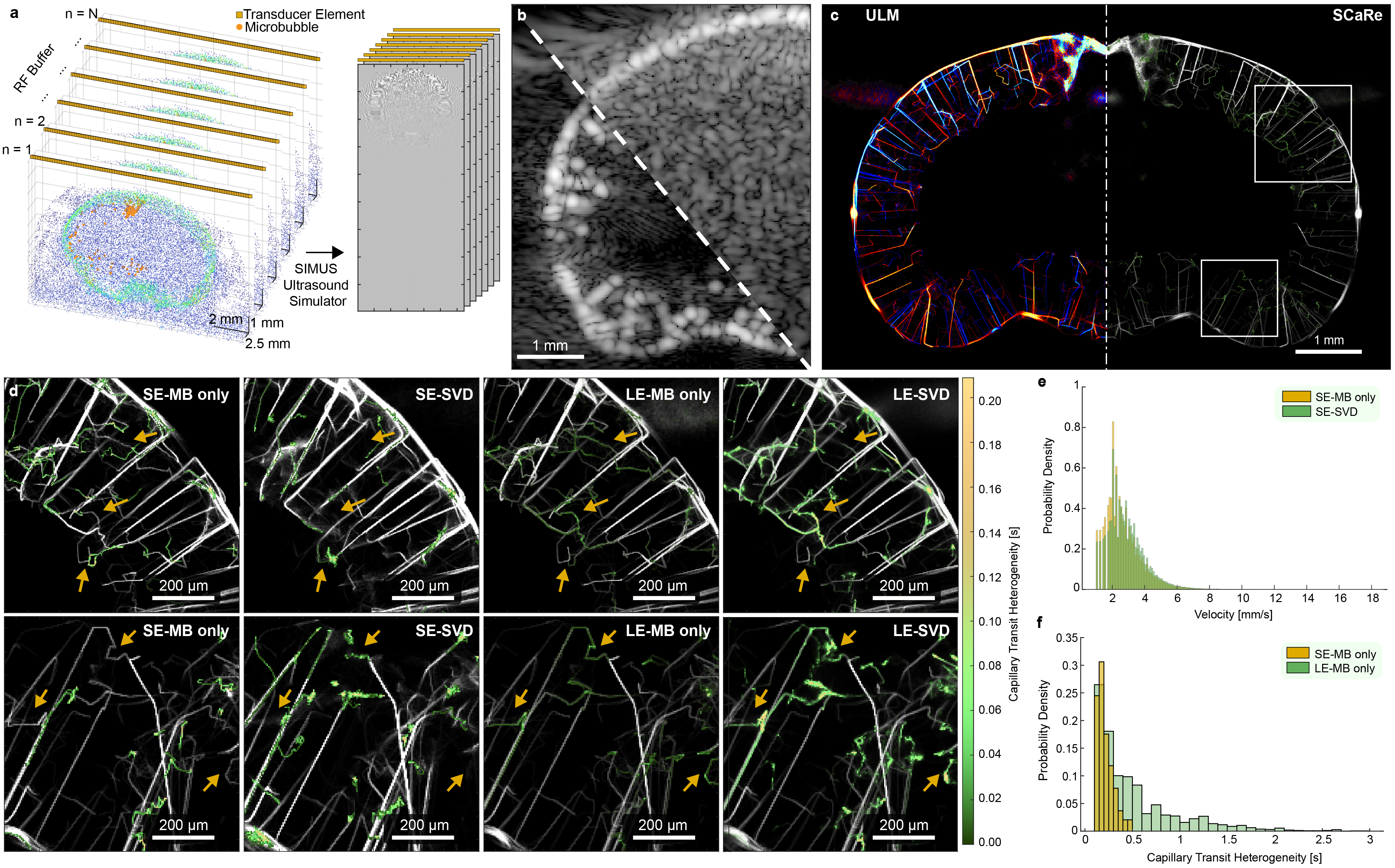}
    \caption{\textbf{a.} Ultrasound data generation pipeline from scatterer distribution \& probe orientation to RF data simulation. \textbf{b.} Representative examples of simulated RF data, beamformed using delay-and-sum, with and without clutter. \textbf{c.} Representative super-resolution ULM-SCaRe maps of accumulated tracks over 350 buffers using long ensemble microbubble only (LE-MB). \textbf{d.} Zoomed images of the ROIs found in \textbf{c.} where the top row corresponds to the top ROI, and the bottom row to the bottom ROI. From left to right: SE-MB only, SE-SVD (20/1000 eigenvalue cutoff), LE-MB only, LE-SVD (20/6000 eigenvalue cutoff) \textbf{e.} Probability density histograms of velocity changes between tracked microbubble-only velocities and SVD filtered velocities in SE conditions. \textbf{f.} Probability density histogram of tracked capillary heterogeneous transit-time, predicted by the trained HMM, for SE vs LE.}
    \label{fig:2}
\end{figure}

We hypothesize that the low velocities found within capillaries poses a problem when using clutter filtering techniques like the SVD. Since the SVD decomposes ensembles of spatiotemporal data to separate strongly decorrelated signals from the uniform skull and brain tissue (located within the first few eigenvalues), it acts like a high pass filter. To investigate the impact of SVD clutter filtering on slow-moving microbubbles, we simulated realistic ultrasound data with added skull clutter on the previously simulated MB dataset. Leveraging an open-source dataset of mouse micro CT scans \cite{FACEBASE:V92}, we generated cluttered skull signals with injected movement to mimic SVD clutter filtering behaviors and artifacts \cite{riemer2023use} (Figure \ref{fig:2}a). These signals were then combined with microbubble centers identified within a 3D coronal slice of the brain, within the footprint of a 16 MHz, 128-element linear transducer. Subsequently, we employed a GPU-accelerated linear ultrasound simulation based on SIMUS equations to generate 350 continuous buffers of ultrasound data, mimicking a continuous \textit{in vivo} acquisition scheme. The resulting signal was beamformed using conventional delay-and-sum techniques (Figure \ref{fig:2}b). Notably, this marks the first instance of simulating combinations of realistic mouse brain and skull clutter that can mimic SVD filtering \textit{in vivo} (see supplementary video 2). Microbubble centers within these signals were successfully localized and tracked using a spatiotemporal tracking method \cite{leconte2023tracking} and accumulated over 350 buffers of ultrasound data to create super resolution maps. We apply our SCaRe methodology to four conditions (short ensemble microbubble SE-MB only, short ensemble SE-SVD filtered, long ensemble microbubble LE-MB only, and long ensemble LE-SVD), where microbubble only indicates no skull clutter. Here, we use all of the tracks located within a ultrasound data buffer to train a Hidden Markov Model to estimate two states: high velocity and low velocity (capillary tracks). Depending on the pattern of the estimated states, we classify whether this track constitutes a capillary or not. The resultant accumulation of tracks allows us to map directional density (figure \ref{fig:2}c left) and map SCaRe overlaid on the vasculature (figure \ref{fig:2}c right).

Analysis of the ground truth, microbubble-only, and microbubble-with-skull-clutter datasets enabled us to investigate the limitations of SVD-based clutter filtering and effects of ensemble sizes. By tracking the microbubble centers over time, we see that SVD filtering significantly hinders the recovery of slow-flowing tracks associated with capillaries. Within the zoomed ROIs (figure \ref{fig:2}d) we see that short ensembles cannot adequately localize capillary behaving microbubbles because the acquisition times (0.5) are too short to sample the whole trajectory. This results in insufficient time to sample the whole U-shaped capillary path and incorrect HMM state estimation. Adding the additional SVD layer results in the same capillary behavior (missing or incomplete tracks shown via arrows). However, the SVD has smoothed some vessels (especially in areas of high concentration near the middle cortex), as well as created new ones, leading to increased false capillaries. Alternatively, if we increase the ensemble size (LE-MB only, LE-SVD), we can better recover the synthetic vasculature, as well as better capillary tracking. We can effectively recover the expected ground-truth capillary tracks in both conditions. We see however, due to SVD filtering, our ability to estimate capillaries, although effective, have lost some detail and resolution.

Furthermore, figure \ref{fig:2}d shows that SVD filtering alters the possible recoverable velocities with fewer slow flow velocities. Compared to SE-MB only, SE-SVD has shifted the velocity distribution to the right, indicating less recovered slow flow velocity tracks. With regards to CTH, figure \ref{fig:2}f illustrates that SCaRe is more effective with tracking over long ensembles with a 500\% increase in max capillary time (Supplementary Figure 1). Given that ULM relies upon the accumulation of thousands of tracks, errors from individual buffers can compound, leading to inadequacies in capillary tracking. Collectively, this experiment demonstrates the feasibility of mapping SCaRe throughout the mouse brain \textit{in silico}, shedding light on the unique characteristics of capillary dynamics as we move towards \textit{in vivo} imaging. 

\subsection{Long Ensemble SVD recovers slow flow microbubbles \textit{in vivo}}\label{subsec3}

Our investigations, so far, reveal the crucial impact of SVD clutter filtering on our ability to recover capillary tracks \textit{in vivo} due to small temporal embedding within the SVD algorithm. i.e., spatiotemporal ensembles of less than 1000 frames formulated as a casorati matrix may embed slow moving microbubbles within the stationary tissue eigenspaces. Specifically, we identify a dual challenge posed by the time required for, and the speed at which, microbubbles traverse through a capillary network. This hinders the recovery of full U-shaped velocity profiles, creating a gap where stationary microbubbles are eliminated (figure \ref{fig:2}f), making clear the need for the acquisition of continuous data to increase ensemble sizes to maximize the likelihood of capturing full capillaries behaviors. Thus, ULM sequences recorded non-continuously for less than 4 seconds may be insufficient to adequately sample CHT (figure \ref{fig:2}f). 

\begin{figure}
    \centering
    \includegraphics[width=0.98\textwidth]{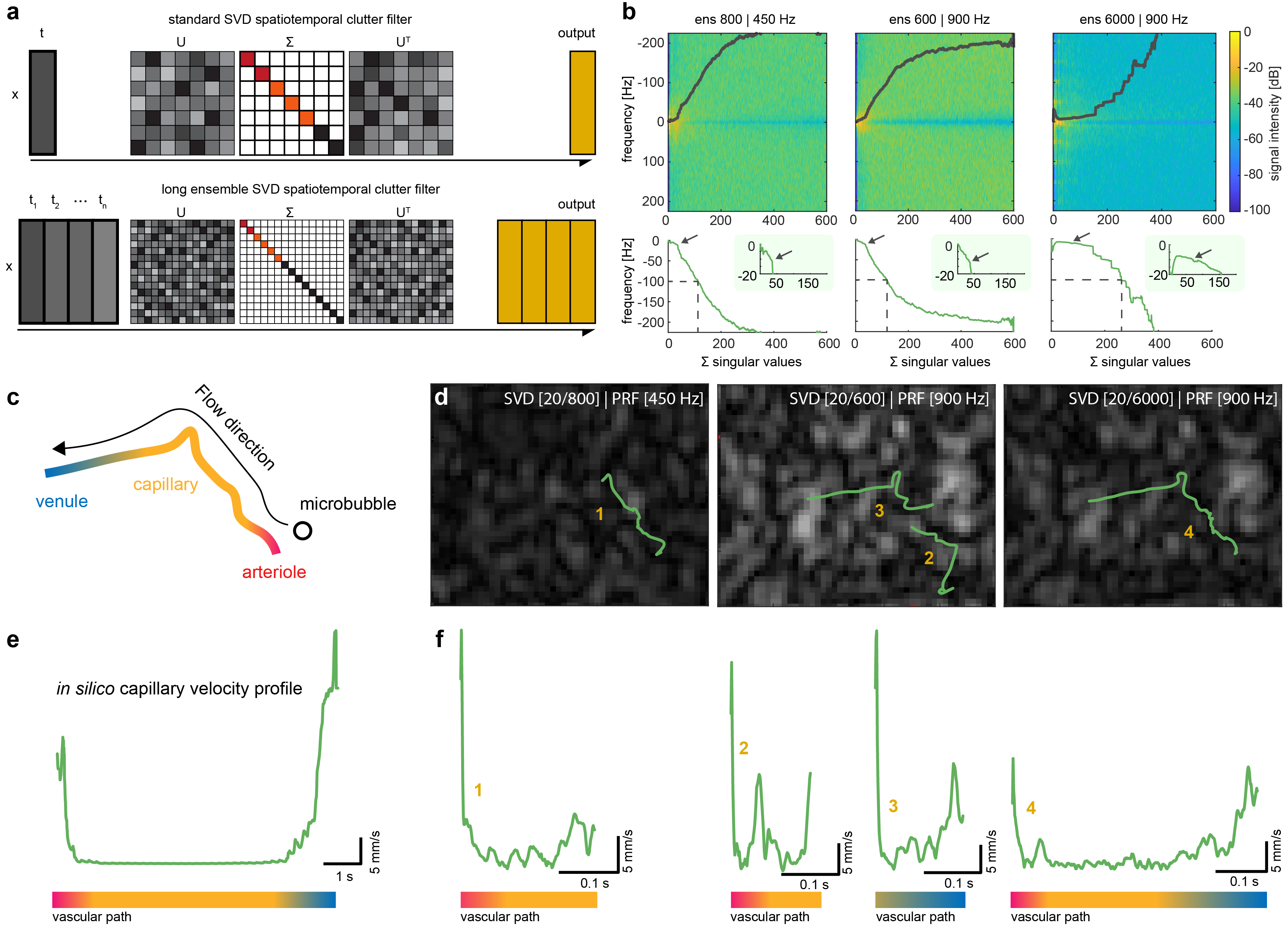}
    \caption{\textbf{a.} Illustration of conventional SVD clutter filtering and LE-SVD methodology. \textbf{b.} Representative power spectrum of 3 different conditions (SE-450 Hz, SE-900 Hz, LE-900 Hz) and their respective mean power spectrum as a function of sorted singular values. Zoom panels are centered around the first inflection point of the spectral density. \textbf{c.} Illustration of an example \textit{in vivo} microbubble flowing through the vascular path from an arteriole through a capillary to a venule. \textbf{d.} representative tracking results with identical parameters over 3 SVD clutter filtering conditions for a single vascular path. The number corresponds to velocity profiles found in \textbf{f}. \textbf{e.} velocity behavior of a microbubble through a whole vascular path in simulated results for comparison to \textbf{f}. \textbf{f.} velocity profiles for \textit{in vivo} tracks of a single microbubble over 3 SVD conditions corresponding to numbers indicated in \textbf{d}.The colorbar in \textbf{e} and \textbf{f} represents the vascular anatomical path in \textbf{c}.}
    \label{fig:3}
\end{figure}

Secondly, the ensemble data size in SVD clutter filtering, crucial for distinguishing slowly moving microbubbles, can be influenced by the frame rate and input to the SVD. Microbubbles traveling at slow speeds, approaching the stationary tissue regime, risk being attenuated in the SVD filtering process if the temporal sampling is inadequate. To understand how to solve this, we apply LE-SVD \textit{in vivo} as a means to retain the full velocity spectrum necessary for recovering U-shaped microbubble velocity behaviors (Figure \ref{fig:3}a).

Here, we compare, on the same dataset, three different conditions \textit{in vivo}, short ensemble SVD (SE-SVD) to LE-SVD, each removing the first 20 eigenvalues: 1) SE-SVD low frame rate (ensemble size 800, 450 Hz), 2) SE-SVD high frame rate (ensemble size 600, 900 Hz), and LS-SVD high frame rate (ensemble size 6000, 900 Hz). Microbubbles were injected via tail-vein catheterization, ultrasound was acquired transcranially through skin, and ULM was constructed using a spatiotemporal tracking methodology \cite{leconte2023tracking}. Under the assumption presented by \citep{Demene2015spatiotemp, 8281060}, the frequency spectrogram as a function of SVD eigenvalues illustrates the subspaces where blood and tissue occupy. Figure \ref{fig:3}b demonstrates spectrograms for the three conditions at the same frequency range (-225 to 225 Hz) and number of singular vectors (1 to 600) where the colormap represents spectrogram intensity. Under the previous assumption, singular vectors whose central frequency (line in Fig. \ref{fig:3}b) is under 100 Hz is thought to be tissue representatives (brain tissue \& skull) and the first inflection point (indicated by arrows) should reveal the change of regime from tissue to blood subspaces. We see for this dataset that SE-SVD conditions (regardless of frame rate) that tissue occupies the subspace below 112 and 119 singular vectors, respectively, whereas in LE-SVD, the tissue subspace is stretched to 262 singular vectors. In the zoom panels in Fig. \ref{fig:3}b, we then see that the inflection points for SE-SVD conditions are 36 and 37 singular vectors, respectively, and at 76 singular vectors in LE-SVD. The ratio between the 100Hz cutoff and inflection points remain consistent (around 30\%) in all cases. Taken together, increasing ensemble size shows that tissue \& skull signals occupy more subspaces where the transition from stationary tissue to uncorrelated blood \& microbubble signal occurs over several singular values, comparable to zero-padding in the frequency domain. Thus, with finer delineation between microbubble and tissue signal, we can maintain the same eigenvalue cutoff with retention of slow flow signal.

Regarding our ability to track single microbubbles through vessels \textit{in vivo} (Figure \ref{fig:3}c), distinct behaviors emerge. Tracking results for three conditions of the same microbubble and vessel in the mouse brain are illustrated in Figure \ref{fig:3}d overlaid on a single B-mode still after SVD filtering. While low frame rates effectively recover slow-moving microbubbles, they fail to capture the fast-moving transition into and out of the capillary. Conversely, high frame rates capture high velocities but miss the MB at low velocities due to insufficient sampling in the SVD, hindering full vessel tracking. In contrast, LE-SVD processed frames produce the entire estimated track, including entry, traverse, and exit from the capillary track. We see that increasing the ensemble size of the SVD can better recover stationary microbubbles (Supplementary Video 3). 

We compare velocity profiles with the ideal U-shaped velocity profile \textit{in silico} (Figure \ref{fig:3}e). The associated velocity profile of the respective figures in Figure \ref{fig:3}d is illustrated in Figure \ref{fig:3}f. While low frame rate SE-SVD only shows the low velocity track, it is prone to false pairing - while high frame rate SE-SVD splits the behavior into two separate tracks. Here, without a priori knowledge, we would not know whether these two tracks are connected. Expectantly, LE-SVD enables recovery of the entire U-shaped velocity profile, exhibiting excellent agreement with \textit{in silico} velocity profiles. We can then combine LE-SVD with track pairing to increases the number of capillary full tracks recovered (Supplementary Figure 1). Thus, these findings demonstrate capability of LE-SVD in overcoming SVD limitations in tracking whole capillary networks \textit{in vivo}.

\subsection{SCaRe biomarkers reveal CHT throughout the whole brain {in vivo}}\label{subsec4}

\begin{figure}
    \centering
    \includegraphics[width = 0.98\textwidth]{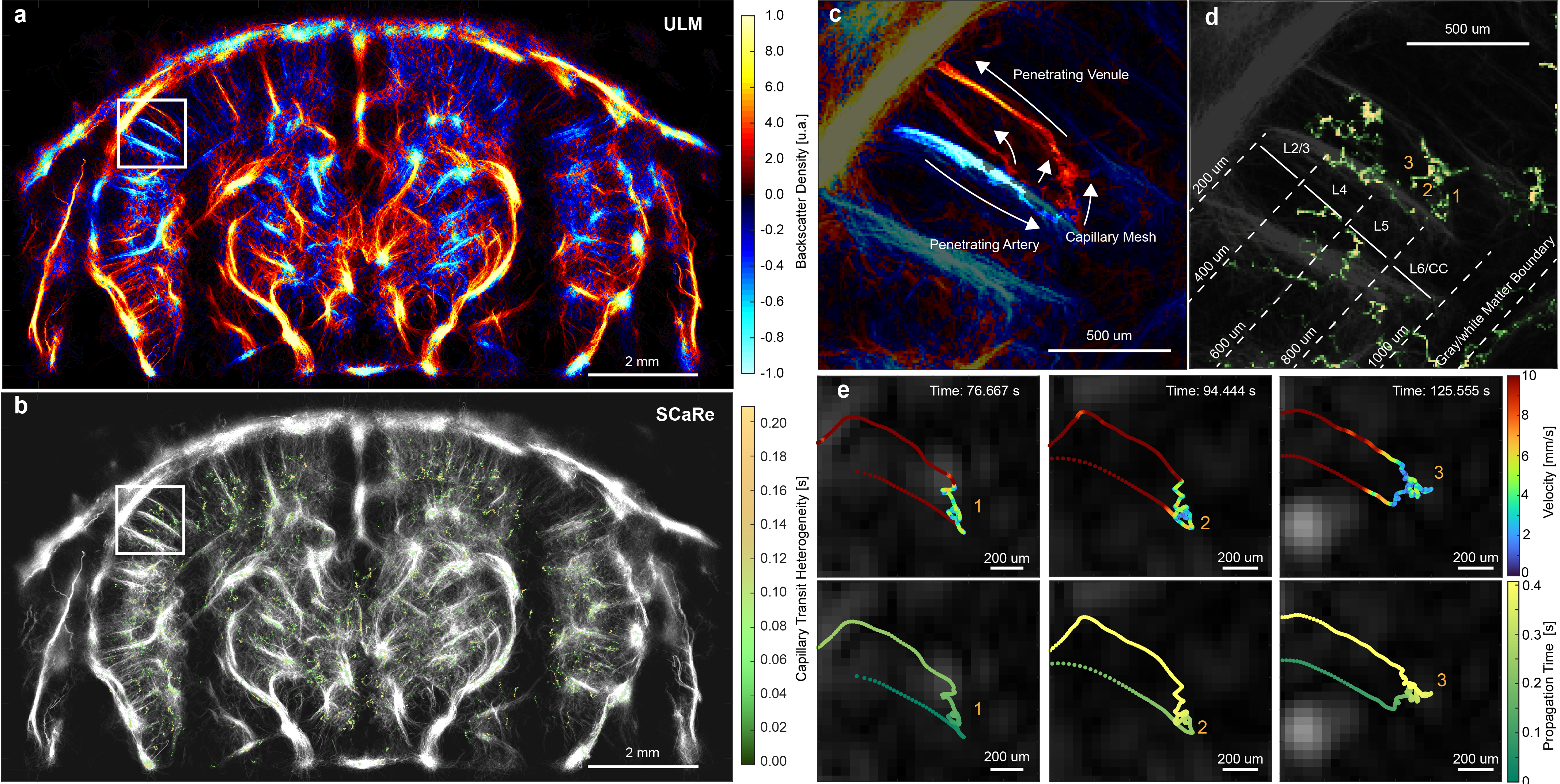}
    \caption{\textbf{a.} Representative ULM maps of a wild-type mouse brain using LE-SVD and track-pairing. \textbf{b.} Representative whole brain SCaRe map of measured capillary tracks with color bar representing the pixel-wise capillary time heterogeneity. \textbf{c.} Zoom panel in \textbf{a} of a completed vascular path with descent in the penetrating arteriole (blue), the ascent out of the penetrating venule (red), and the capillary mesh in between. \textbf{d.} Zoom panel in \textbf{b} illustrating tracked capillary networks in neuronal layers 5 and 6. \textbf{e.} tracks from 3 different capillary paths through the same arteriole and venule separated by microbubble velocity and time of travel through the vascular network.}
    \label{fig:4}
\end{figure}

Here, we comprehensively map whole-brain SCaRe \textit{in vivo} via continuous data acquisition of a bolus of Definity microbubbles (1:10 dilution), enhanced clutter filtering using LE-SVD, and the implementation of HMMs trained to discern MB descent and ascent through capillaries. Figure \ref{fig:4}a illustrates a track density map using backscattering amplitude for display and figure \ref{fig:4}b shows the corresponding SCaRe map (Bregma = -1.5 mm). Notably, distinct single capillary tracks are discernible in both cortical and subcortical regions of the brain. With this strategy, we can identify single capillaries in the zoomed panels in Figure \ref{fig:4}c to isolate single vascular paths composed of a penetrating arteriole connected to penetrating venules. Remarkably, this map illustrates a composite of four separate MB paths originating from the same arteriole (blue; descent) with one leading to a separate venule (red; ascent). The ability to discern capillaries in this fashion facilitates quantification of heterogeneity in the spatiotemporal patterns of capillaries in ULM, which was previously challenging using the existing techniques.

Furthermore, Figure \ref{fig:4}d showcases the ability to segregate measured CHT by neuronal layers. This capability is particularly pertinent given previous studies indicating layer-specific differences in capillary function, notably in age-related white matter loss \cite{stamenkovic2024impaired}. Individual SCaRes can be analyzed to discern their travel path, velocity behaviors, and estimated CHT, as depicted in Figure \ref{fig:4}d. As expected from our simulations, capillaries identified using SCaRe exhibit high-velocity entry, followed by low-velocity capillary transit, and culminating in high-velocity uptake into penetrating venules. Notably, figure \ref{fig:4}e illustrates velocity and propagation time for these specific capillaries, tracked over an entire minute (supplementary video 4). Here, we can identify exactly when the microbubble enters a region of low velocity (\(\textless\) 2 mm/s) with differences in the microbubble travel path within the capillary mesh. The propagation time shows that these microbubbles enter from the same arteriole and exit from the same venule with varying transit-times. The observation of differing capillary transits originating from the same arteriole and exiting through the same venule underscores the spatiotemporal complexity of microvascular dynamics. With SCaRe, the capability to elucidate such patterns for single capillaries presents numerous avenues for advancing our understanding of brain health, such as how the brain ages as well as how neuronal function is impacted by capillary heterogeneity.

\begin{figure}
    \centering
    \includegraphics[width = 0.98\textwidth]{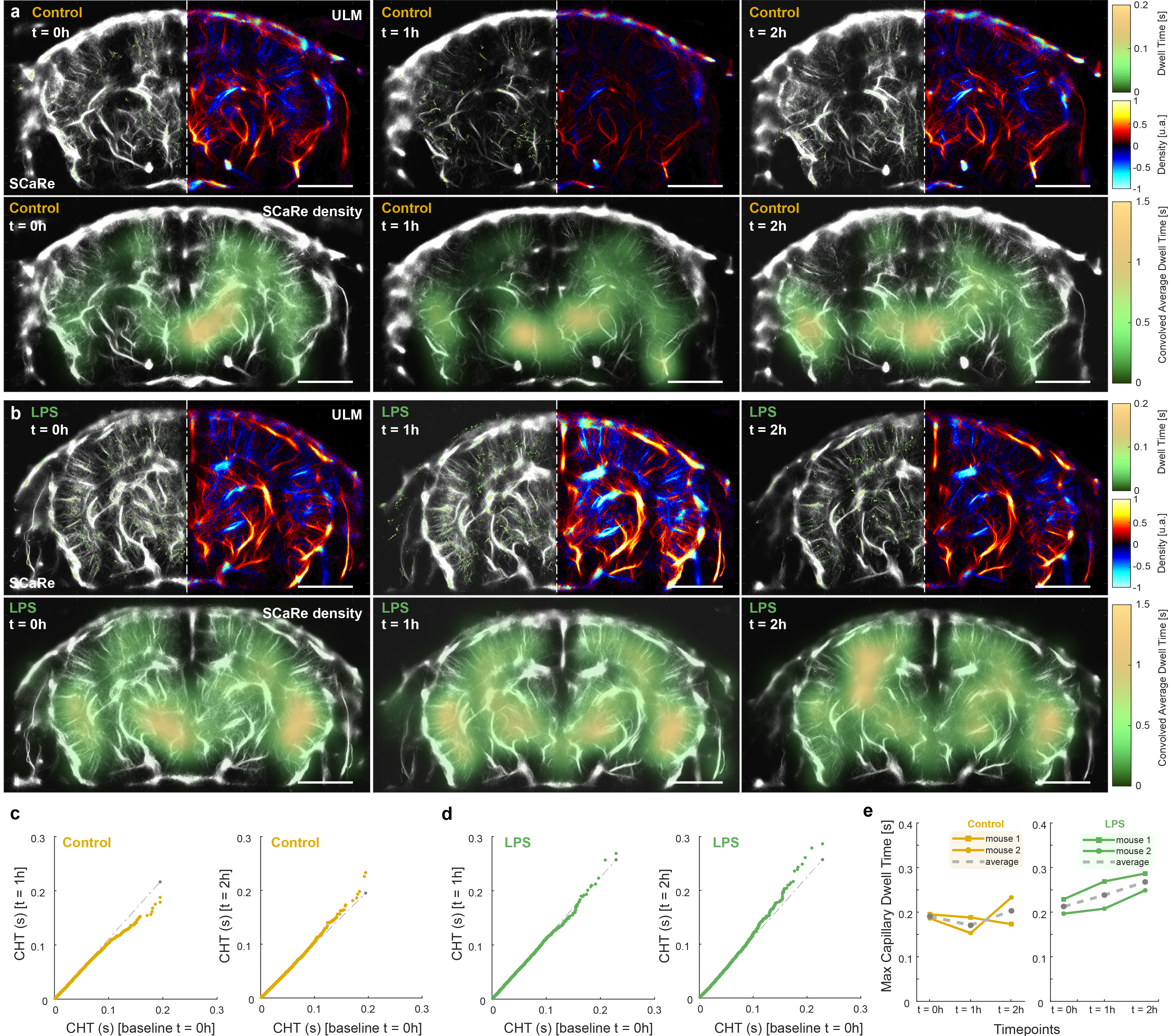}
    \caption{\textbf{a.} Representative SCaRe ULM at time points 0h, 1h, and 2h for negative control mice (no injection). \textbf{b.} Representative SCaRe ULM at time points 0h, 1h, and 2h after LPS injected mice. \textbf{c.} QQ plots of CHT for control mice comparing time points 1h and 2h against the baseline (0h). \textbf{d.} QQ plots of CHT for LPS injected mice comparing time points 1h and 2h against the baseline (0h). \textbf{e.} Max dwell time over the entire 5 minute scan time (n = 2 per group) as a function of time points for control and LPS mice. The mean value is shown as the grey dotted line.}
    \label{fig:5}
\end{figure}

Lastly, we asked whether SCaRe biomarkers can detect sensitive changes to capillary function throughout the whole brain. As we understand, intraperitoneal (IP) injection of lipopolysaccharide (LPS) causes systemic inflammation, increasing recruitment of leukocytes \cite{qin2007systemic}. But as a consequence, leukocytes plug small cerebral capillaries, increasing the blood propagation time not only in the micro-stalled capillary, but increasing CHT globally \cite{jamshidi2024impact}. Thus, to  validate SCaRe, we measured the CHT - within the same animal and imaging plane (n = 2 control, n = 2 LPS; 5.5 month old) - before and after injection of LPS with comparison to a negative control (without IP LPS injection). Figure \ref{fig:5}a illustrates representative whole brain images of the negative control group at 3 time points (0h - baseline, 1h, and 2h after LPS injection). SCaRe and ULM maps are shown, as well as a SCaRe-density map (SCaRe convolved with a Gaussian kernel), for each time point. Figure \ref{fig:5}b shows matching SCaRe, ULM, and SCaRe-density brain maps of LPS injected mice. Qualitatively, both control mice SCaRe and SCaRe-density show similar levels of CHT across time points with small decrease shown in the cortex. Conversely, SCaRe maps in LPS injected mice show monotonic increase, especially in the left hemisphere of the cortex.

To compare CHT distributions across the 5 minute scan, we construct QQ-plots of 1h and 2h post-injection time points against the baseline 0h (Figures \ref{fig:5}c-d). Here, probability quantiles are compared such that identical distributions lie along the 45$^\circ$ line (dotted gray line). As LPS systemic-inflammation is expected to increase capillary transit time \cite{qin2007systemic}, we expect to see increases in the distribution tails such that QQ plots diverge from the 45$^\circ$ line away from the baseline. In control mice, we see small oscillation at the tails (small decreases at t = 1h and increases at t = 2h) with similar distributions at 0.1 s CHT. These fluctuations are likely due to biological factors, such as the anesthetic agent used. However, in LPS injected mice we see increases in the tails at both time points with more significant increase at t = 2h post-LPS injection. This data suggests that SCaRe tracked more capillaries that had longer transit times. Moreover, the longer range of illustrated distributions in Figure \ref{fig:5}d indicate that the SCaRe-measured capillary transits were higher in LPS-injected mice. In fact, if we measure the maximum CHT value throughout the whole 5 minute scan in both populations, we see a monotonic 25.9\% total increase in LPS mice that we do not see in the negative controls. Statistical testing using a mixed linear model regression analysis give some indication that bolsters our hypothesis (Supplementary table 1). Firstly, there is significance in max CHT values at time points, 1h (p = 0.032) and 2h (p = 0.03) post-injection in the LPS group with no difference between baseline. The interaction parameter between time and group is significant for t = 1h, and marginally not significant for t = 2h, indicating differences in the max CHT are due to effects of LPS injections over time. Taken together, our results show a clear connection from quantified SCaRe measurements to capillary function.

\section{Discussion}\label{sec3}

In this study, we have demonstrated the efficacy of SCaRe in identifying and discriminating singular capillary tracks within ULM images. By employing realistic computational models of the mouse brain with fully-connected vascular trees, we have uncovered specific limitations in ULM, including the requisite acquisition length and the challenges in distinguishing slow-moving microbubbles. Furthermore, our investigation has revealed distinctive U-shaped velocity profiles exhibited by microbubbles traversing capillary vessels. Leveraging a combination of LE-SVD, spatiotemporal tracking, track pairing, and HMMs, we have successfully addressed these inherent limitations and accurately identified single capillaries amidst a forest of tracks. Applying these strategies to \textit{in vivo} imaging of mouse brains after LPS injections, we have confirmed the feasibility of SCaRe, identifying lower velocities and the unique U-shaped velocity profile of microbubbles transitioning from arterioles to venules. Subsequently, through minute-long scans, we have mapped capillary function throughout the entire brain at depth, a feat unachievable by existing optical, ultrasound, CT, or MRI-based neuroimaging modalities.

The implications of SCaRe yield new insights into observing the brain both under normal function and neurological conditions. In this study, we use SCaRe as a correlate for capillary transit time. However, it opens avenues for exploring other potential biomarkers, such as investigating heterogeneity within a single capillary mesh (Figure \ref{fig:4}\textbf{c-e}) or understanding local capillary function within specific brain regions or cortical layers \cite{stamenkovic2024impaired}. While the measured CTH \textit{in vivo} (1.4s) falls below the maximum CTH (4.0s) observed in simulations, our measurements appear consistent with those reported in the literature \cite{gutierrez2016effect} under steady-state conditions. Furthermore, the use of LPS demonstrates the feasibility of SCaRe in detecting subtle changes within CTH distributions, as increasing capillary stalls result in flow variations and increased arteriovenous transit times, as measured by optical methods \cite{jamshidi2024impact}. Although LE-SVD enables us to recover smaller flow rates and trace more tracks compared to conventional ULM, discrepancies between simulation and \textit{in vivo} CTH measurements may still arise due to the difficulty in tracking stationary microbubbles. Though SCaRe may be computed using other tracking algorithms (Gaussian localization with nearest-neighbors, or Hungarian pairing), the use of a spatiotemporal tracking method, where localization and tracking are interdependent, exerts higher confidence in retaining the microbubble behavior. Moreover, the influence of the SVD ablation of slow moving microbubbles may lead to larger downstream effects in independent localization and tracking algorithms. Nevertheless, capillary stalls and occluded capillaries may exert systemic effects throughout the microvasculature, suggesting potential avenues for further inference and analysis.

At its core, SCaRe relies on the extraction of U-shaped velocity profiles, depicting the high-velocity descent through arterioles, slow perfusion through capillaries, and high-velocity uptake into venules. While reminiscent of sensing-ULM (sULM) \cite{denis2023sensing,chabouh2024whole} used to identify circling microbubbles (\(\textless\) 1s) within glomeruli in the kidney, our methodology differs significantly due to the distinct behavior of brain capillaries and challenges in transcranial imaging. Our approach remains agnostic, solely dependent on the estimated velocity profile, which may prove useful for identifying glomeruli as well. However, a major limitation of SCaRe is its dependence on accurately tracking a microbubble over seconds, highlighting the importance of preceding algorithms involved in clutter filtering, localization, and tracking. The incorporation of LE-SVD has shown promise in recovering slower microbubble flows, enhancing our ability to spatiotemporally pair tracks and delineate the complete vascular travel path. Additionally, though SCaRe now can tell us whether a single track is a capillary, it lacks structural information on vessel diameter since that requires the presence of multiple tracks to confirm vessel morphology. %Given more complex synthetic hemodynamic models, we may be able to create models that predict capillary size given the size of connected arterioles and venules. 

Further considerations must be given to implementing SCaRe effectively. Continuous acquisition is paramount for tracking microbubbles over seconds, necessitating careful adjustment of pulse repetition frequency (PRF) and emission pressure to avoid bubble bursting. Our study has demonstrated successful capillary recovery using specific PRF ranges, yet optimization is needed to balance bubble stability and data acquisition efficiency, especially given the need for recovering tracks that include both high and low velocity regions. Additionally, the choice of transducer and driving voltage (i.e., derated pressures) presents challenges, particularly in determining the ideal parameters for skull thickness variations in procedure.

Other limitations arise from using a 2D linear array to capture inherently 3D information. While this limitation extends beyond ULM imaging, future exploration into 3D imaging techniques may unlock more complex capillary hemodynamics and reveal spatiotemporal relationships between brain regions which may be available for exploration with the simulation tool we developed here. While our 3D computational model has relatively simple capillary connections compared to \textit{in vivo} 2-photon images, it provides essential groundwork for SCaRe development. More studies, however, need to be performed in order to acquiesce the ideal concentrations and scan times to adequately sample the capillary transit-time. Our \textit{in vivo} experiments indicate concentration highly affects our ability to recover SCaRe with an inverse relationship to the injected microbubble bolus, where more SCaRe is recovered after the initial increase. This challenge may be overcome through the use of controlled continuous perfusion.

Lastly, we show the feasibility of using statistical methods, such as Hidden Markov Models, to characterize the behavior of our tracked microbubbles. However, it is known that HMMs are simplistic - assuming independence between observed states -, are prone to over or under fitting, and are sensitive to initial parameters, as well as the quality of the training data. Here, we model the problem as a 2-state HMM and indeed, the observed velocity traces of SCaRe indicate successfully retrieve the expected U-shaped profile. However, this methodology will benefit from more sophisticated implementation of HMMs, such as those found in cell or particle tracking \cite{das2009hidden}, or data-driven deep learning solutions \cite{nehme2020deepstorm3d}. 

Future studies will explore different microbubble behavior classifications to address computational challenges associated with HMMs and investigate implementing skull aberration correction. As induced systemic neuroinflammation demonstrated the ability to measure subtle changes in SCaRe, additional efforts will focus on identifying functional spatiotemporal relationships in SCaRe across various healthy and disease states, potentially aiding in stroke monitoring and capillary stalling detection. Moreover, the methodology developed through SCaRe begets other possible biomarkers, such as the interdependence of capillary function over neuronal layers, as well as the velocity heterogeneity tracked within a single capillary. Understanding capillary function with ULM may have large implications for how we treat and care for Alzheimer's and stroke patients. Lastly, the versatility of SCaRe extends beyond cerebral microvascular imaging, potentially benefiting all vascularized organs, contrast-based imaging methods, and localization and tracking domains.

\section{Methods}\label{sec4}

\textbf{Simulations} 
\textit{In silico} simulations were built on top of a computational model of mouse brain microvasculature with fully connected capillary networks \cite{linninger2019mathematical}. These models were interfaced with MATLAB's graph and network algorithms module and contained information pertaining to vessel size, blood flow magnitude and direction, and pulse pressure. From here, sequential Monte Carlo simulations were built on the vascular network to generate datasets of microbubble flow through the mouse cortical networks \cite{belgharbi2023anatomically}. Specifically, we randomly sampled from a simulated distribution of modeled Definity microbubbles (mu = 2 um, std = 3 um) \cite{talu2007tailoring}. For one hemisphere, there are 291,372 possible paths through capillaries. Thus, each microbubble was forward simulated taking into account the size of the bubble, the flow velocity (calculated from instantaneous blood flow and vessel diameter), as well as the pulse wave velocity, simulated as a traveling wave of 350 cm/s at 500 BPM \cite{marshall2023alterations,janssen2016need}. Additional constraints were put so that a bubble cannot pass through a capillary smaller than the diameter. Afterwards, microbubble datasets were stitched together, syncronous with the cardiac cycle, to produce a fully populated dataset of microbubbles moving through the mouse cerebral microvasculature.

To further simulate ultrasound data from the microbubble dataset, we aligned the microvasculature with an open source repository for mouse micro CT \cite{FACEBASE:V92}. Then, a 3D portion of the microbubbles and skull micro CT were segmented to fit within the elevational and lateral footprint of a 16 MHz 128-element probe (spacing = 100 $\mu m$, element height = 1.5 mm). 12 scatterers per resolution cell ($\frac{\lambda}{4}$) were simulated and arranged according to the intensity of the $\mu$CT so that the scatterers placed at the skull were of higher concentration than the brain matter. These scatterers were superimposed with the centers of the microbubbles for input into a linear ultrasound simulator. A GPU-accelerated simulator based on the equations from SIMUS \cite{cigier2022simus} was used to simulate IQ data for 3500 plane waves (7 angles between $-11^\circ$ and $11^\circ$) at 7000 Hz PRF. Simulated results were stored so that the full IQ datasets were linearly superimposed RF from microbubbles and RF from skull clutter. 

\noindent\textbf{Animals.}
All animal experiments were performed in accordance with the Animal Research Ethics Committee of the Montreal Heart Institute (Protocol \# 2023-32-02 TAC-ultrasons).  C57BL/6 wild-type mice were employed in this study, (n = 2 male, n = 2 female, 8 - 10 weeks old). For LPS experiments n = 4 mice, 5.5 months old, were used, n = 2 negative control and 2 LPS injection validation. Animals were anesthetized with isoflurane (2\%, 1L O$_2$ induction; 1-1.5\%, 0.5-1L O$_2$ management). After induction, mice were secured in a sterotaxic frame (SGM-4, Narishige), dehaired at the tops of the heads, and catheterized for tail-vein injections using 25G needles. Physiological saline was used to test vessel patency and confirm tail-vein placement. Degassed ultrasound gel was placed on top of the head, and a warm-water bath was placed immediately on top to couple the transducer to the head.

Microbubble bolus injections were prepared using 4 uL/g of body weight of Definity microbubbles (Lantheus, MA, USA), diluted in physiological saline at a 1:10 ratio. One bolus was used, up to a max of 4, for each ultrasound scan lasting 3 - 5 minutes. The injection of the bolus was immediately followed by a 50-60 uL sterile saline wash. Systemic inflammation was triggered by lipopolysaccharide (LPS) injections purified from E. coli strain O111:B4 (Sigma-Aldrich, St. Louis, USA) at a single dose of 2mg/kg in 200µL saline. LPS injections were applied intraperitoneal after isoflurane induction and an initial bolus scan for baseline SCaRe mapping. Additional bolus' and scanning was performed at 1 and 2 hours post LPS injection to assess changes in CTH.

\noindent\textbf{Ultrasound Acquisition.}
All ultrasound data was acquired transcranially through skin using a linear hockey-stick array with a central frequency of 10.4667 MHz and a bandwidth between 8 and 18 MHz (L8-18iD, GE, IL, USA) and interfaced with a 256 Vantage research ultrasound machine (Verasonics, WAS, USA). IQ data was acquired at 50\% bandwidth sampling. Seven tilted plane waves (-5$^\circ$ to -5$^\circ$) at a compounded frame rate (900 Hz - 1700 Hz) for a total of 8100 and 15300 Hz PRF were acquired for 1 buffer. Ultrasonic parameters were chosen to be less than or equal to the transfer time of 4 buffers to an NVMe SSD so that data can be continuously acquired and streamed. A single ULM scan was taken between 3 and 5 minutes, acquiring 350 to 400 GB of data. Offline, all IQ data was beamformed onto a standard grid with spacing of $\lambda/2\sqrt{2}$ using standard GPU-based delay-and-sum beamforming.

\noindent\textbf{Ultrasound Localization Microscopy.} 
ULM tracks were processed using a sliding window of 4 buffers on 6 stacks of buffers (6000 total frames). IQ data underwent processing by first removing the mean and standard deviation in slow time, followed by application of conventional short ensemble SVD (SE-SVD), SE-SVD stitched, or long ensemble SVD (LE-SVD). All SVDs were computed using the casorati formulation and matrix decomposing the covariance where increased ensembles increases the computational time. The power frequency spectrum was computed via a periodogram with a 0.2 tukey apodization sorted by decreasing eigenvalues \cite{Demene2015spatiotemp}. For each condition, the first 20 eigenvalues were eliminated to remove stationary tissue signal from the skull and brain tissue, leaving only the blood and microbubble signals. 

Subsequently, a spatiotemporal tracking algorithm was employed to track microbubbles in space and time \cite{leconte2023tracking}. This approach facilitated direct access to the spatiotemporal signal and velocity of the track, which was utilized for HMM categorization. Modifications were made to the vesselness filter to account for the temporal component in constructing vessel profiles, and the spatiotemporal tracker was adjusted to include track pairing. Lastly, backscattering amplitude \cite{renaudin2023backscattering} was recorded for each track as the absolute value of the IQ and the mean value of all the tracks per pixel were subsequently used for display on ULM and SCaRe maps. 

Initially, the starts and ends of each track were paired if they were equivalent points in space and time. Subsequently, a second track pairing was performed to chronologically pair tracks with a distance (in x and z) less than 1$\lambda$ with a temporal gap of less than 100 ms. Additionally, a Kalman filter was implemented immediately after radial symmetry localization. The Kalman filter was initialized with the first coordinate of the track and updated using the following equations:

\textbf{Prediction:}
\[
\begin{aligned}
x_{\text{pred}} &= F \times x \\
P_{\text{pred}} &= F \times P \times F^T + Q
\end{aligned}
\]

where \( F \) is the state transition matrix, \( P \) is the identity matrix, and \( Q \) is the process noise covariance \(\sigma\). \(\sigma\) was set empirically. These matrices are initialized as:
\[ 
    F = \begin{bmatrix} 1 & 0 & \text{dt} & 0 \\ 
        0 & 1 & 0 & \text{dt} \\ 
        0 & 0 & 1 & 0 \\ 
        0 & 0 & 0 & 1 \end{bmatrix},
    P = \begin{bmatrix} 
        1 & 0 & 0 & 0 \\ 
        0 & 1 & 0 & 0 \\ 
        0 & 0 & 1 & 0 \\ 
        0 & 0 & 0 & 1 \end{bmatrix},
    Q = \begin{bmatrix} 
        \sigma & 0 & 0 & 0 \\ 
        0 & \sigma & 0 & 0 \\ 
        0 & 0 & \sigma & 0 \\ 
        0 & 0 & 0 & \sigma \end{bmatrix} 
\]

\textbf{update:}
\[
\begin{aligned}
    y &= \text{nint}\left(\frac{z - H \times x}{g}\right) \times g \\
    S &= H \times P \times H^T + R \\
    K &= \frac{P \times H^T}{S} \\
    x_{\text{pred}} &= x + K \times y \\
    P_{\text{pred}} &= (I - K \times H) \times P \\
\end{aligned}
\]
Where z is actual 2D measurement, g is the desired grid spacing for normalizing the discrete values output by the centerline thinning algorithm, H is the measurement matrix, R, is the measurement noise $\epsilon$ covariance, S is the residual covariance, and K is the Kalman gain. H and R are initialized as:
\[ 
    H = \begin{bmatrix} 1 & 0 & 0 & 0 \\ 
        0 & 1 & 0 & 0   \end{bmatrix},
    R = \begin{bmatrix} 
        \epsilon & 0 \\ 
        0 & \epsilon \end{bmatrix}
\]

After the initial pass through prediction and update, a RTS smoothing filter \cite{rauch1965maximum} was applied in reverse according to the following equations:

\[
\begin{aligned}
C &= \frac{P_{\text{pred}}(t) \times F^T}{F \times P_{\text{pred}}(t+1) \times F^T + Q} \\
\hat{x}(t) &= x_{\text{pred}}(t) + C \times \hat{x}(t+1) - F \times x_{\text{pred}}(t) \\
\hat{P}(t) &= P_{\text{pred}}(t) + C \times (\hat{P}(t+1) - F \times P_{\text{pred}}(t)) \times C^T \\
\end{aligned}
\]

Where C is the smoother gain, \(\hat{x}\) and \(\hat{P}\) are the updated state and covariance after smoothing. Capillary HMM training was conducted at this stage in the pipeline. Subsequently, tracks were spline interpolated and derivated for velocity display on ULM maps. A direction filter was applied to tracks to differentiate the colormap for upward movement (hot) and downward movement (cold) based on the sign of the velocity in the z direction. Finally, these tracks were interpolated and accumulated on a super-resolved grid of $\lambda / 32$, and the velocities were displayed over the track density map.

\noindent\textbf{SCaRe processing.}
SCaRe processing encompasses three main stages: HMM training, state prediction, and capillary categorization. Initially, all tracks undergo preprocessing to convert velocity profiles into observations comprising four states based on velocity and acceleration. Subsequently, all observations are standardized to the same length to facilitate Hidden Markov Model (HMM) training using the Baum-Welch algorithm implemented in MATLAB's hmmtrain function. The initial transmission probabilities are represented by a 2 x 2 matrix, while the emission probabilities are structured as a [2 x 4] matrix.

In the second stage, state predictions on tracks are generated using the Viterbi algorithm (hmmviterbi), estimating whether the profile corresponds to a high-velocity state or a low-velocity state indicative of capillary flow.

Finally, in the third stage, these states are characterized based on the occurrence and location of low-velocity states. Specifically, we identify whether the capillary network exhibits a U-shaped profile with a minimum capillary heterogeneous transit time threshold of 0.120 s. Here, Capillary Heterogeneous Transit Time is defined as the duration spent in the low-velocity state between the inlet and outlet points.

For visualization purposes, capillary trajectories are processed by accumulating centered positions within the HMM viterbi segmented regions of the super localized track positions. The subsequent map thus shows the integral over time given by the frame rate, highlighting pixels where microbubbles remain stationary, indicating time spent in the capillary. Subsequently, SCaRe maps are superimposed onto grey-scale ULM maps of all trajectories, effectively integrating functional and structural information for comprehensive visualization and analysis. To illustrate more global capillary information, and the SCaRe map was convolved using a $8\lambda$ window Gaussian.

\noindent\textbf{Statistical Testing.}
Distributions of CHT in negative control (n = 2) and LPS-injected mice (n = 2) for the entire 5 minute scan (at 3 time points: 0h-baseline, 1h-post, and 2h-post) were qualitatively compared using QQ-plots (qqplot in MATLAB). These CHT were acquired from the SCaRe map and only tracks within a manually segmented ROI of the brain tissue were used for analysis. To quantitatively compare max CHT acquired at all three time points and their interactions between groups, we used a Mixed Linear Model Regression algorithm (mixedlm, part of the statsmodels python package).

\backmatter

\bmhead{Acknowledgments}

This work was supported by the Vanier-Banting Postdoctoral Fellowship from the National Science, Engineering, and Research Council (NSERC), the TransMedTech Living Lab Postdoctoral Fellowship, the Institute for Data Valorization (IVADO), in part by the Canada Foundation for Innovation under Grant 38095, in part by the Canadian Institutes of Health Research (CIHR) under Grant 452530, in part by the New Frontiers in Research Fund under Grant NFRFE-2018-01312 and in part by the Natural Sciences and Engineering Research Council of Canada (NSERC) under Grant RGPIN-2019-04982.  Further support came from the Fonds de recherche du Québec - Nature et technologies, the Quebec Bio-Imaging Network, the CONAHCYT, the Insightec, the Healthy Brains Healthy Lives, and the Canada First Research Excellence Fund. Additionally, computational resources were provided through the Digital Research Alliance of Canada. A.Linninger acknowledges funding by NIH National Institute of Aging (NIA) 1R01AG079894

\section*{Declarations}

\subsection{Author's Contributions}
S.A.L and J.Pr conceived the study. A.Li developed the realistic mouse cerebral microvasculature dataset and the graph based methodology for hemodynamic simulations of blood flow and pressures. S.A.L and J.K developed the microbubble Monte Carlo simulations. S.A.L, J.Po developed the ultrasound simulation process. S.A.L and A.Le developed the ULM data-processing algorithms. S.A.L and A.W acquired the \textit{in vivo} data. S.A.L developed the SCaRe methodology and algorithms. S.A.L wrote the first draft of the manuscript with substantial contribution from J.Po and J.Pr. All authors edited and approved the final version of the manuscript.

\subsection{Competing interest}
The authors have no competing interests to disclose.

\subsection{Data availability}
The data supporting the findings of this study are provided within the paper and its Supplementary Material. The raw and analyzed datasets generated, such as the microvascular model, the microbubble model, the ultrasound simulated data, and \textit{in vivo} datasets are available for research purposes from the corresponding author upon reasonable request.

\subsection{Code availability}
The SCaRe algorithims are available upon reasonable request from the corresponding author.

%%===========================================================================================%%
%% If you are submitting to one of the Nature Portfolio journals, using the eJP submission   %%
%% system, please include the references within the manuscript file itself. You may do this  %%
%% by copying the reference list from your .bbl file, paste it into the main manuscript .tex %%
%% file, and delete the associated \verb+\bibliography+ commands.                            %%
%%===========================================================================================%%

\bibliography{sn-bibliography}% common bib file
%% if required, the content of .bbl file can be included here once bbl is generated
%%\input sn-article.bbl

\end{document}